\documentclass[conference,10pt]{IEEEtran}
\IEEEoverridecommandlockouts
\usepackage{cite}
\usepackage{amsmath,amssymb,amsfonts}
\usepackage[ruled,lined]{algorithm2e}
\usepackage{algpseudocode}
\usepackage{graphicx}
\usepackage{textcomp}
\usepackage{xcolor}
\usepackage{lettrine}
\usepackage{multirow}
\usepackage{caption}
\usepackage{subcaption}
\usepackage{tabularx}
\usepackage{romannum}
\usepackage{array}
\usepackage{cleveref}

\newcommand{\bi}{\begin{itemize}}
\newcommand{\ei}{\end{itemize}}
\newcommand{\be}{\begin{enumerate}}
\newcommand{\ee}{\end{enumerate}}
\newcommand{\bd}{\begin{description}}
\newcommand{\ed}{\end{description}}
\newcommand{\bc}{\begin{center}}
\newcommand{\ec}{\end{center}}
\newcommand{\bt}{\begin{tabbing}}
\newcommand{\et}{\end{tabbing}}

\newcommand{\bfig}{\begin{figure}}
\newcommand{\efig}{\end{figure}}
\newcommand{\beq}{\begin{equation}}
\newcommand{\beqarr}{\begin{eqnarray}}
\newcommand{\beqarrn}{\begin{eqnarray*}}
\newcommand{\eeq}{\end{equation}}
\newcommand{\eeqarr}{\end{eqnarray}}
\newcommand{\eeqarrn}{\end{eqnarray*}}
\newcommand{\bflr}{\begin{flushright}\vspace{-0.2in}}
\newcommand{\eflr}{\end{flushright}}
\newcommand{\bsub}{\begin{subequations}}
\newcommand{\esub}{\end{subequations}}
\newcommand{\barr}{\begin{array}}
\newcommand{\earr}{\end{array}}

\newcommand{\vect}[1]{\boldsymbol{#1}}
\newcommand{\norm}[1]{\left\|#1\right\|}

\usepackage[textwidth=30mm]{todonotes}

\newcommand{\hrulealg}[0]{\vspace{1mm} \hrule \vspace{1mm}}
\newcommand*{\rom}[1]{\expandafter\@slowromancap\romannumeral #1@}

\SetKwFor{For}{for}{do}{endfor}
\begin{document}


\title{Adversarial Robustness of Distilled and Pruned Deep Learning-based Wireless Classifiers \vspace*{-0.5em}}

\author{Nayan Moni Baishya and B. R. Manoj \\
Department of Electronics \& Electrical Engineering, 
Indian Institute of Technology Guwahati, India.\\
Emails: {\tt{\{nmb94, manojbr\}}@iitg.ac.in}
\vspace*{-1.25em}
\thanks{This work was supported in part by SERB Start-Up Research Grant (SRG) Scheme, Department of Science and Technology (DST), Govt. of India under Grant SRG/2022/001214 and in part by Start-Up Grant of Indian Institute of Technology Guwahati.
} 
}

\maketitle

\begin{abstract}
Data-driven deep learning (DL) techniques developed for automatic modulation classification (AMC) of wireless signals are vulnerable to adversarial attacks. This poses a severe security threat to the DL-based wireless systems, specifically for edge applications of AMC. In this work, we address the joint problem of developing optimized DL models that are also robust against adversarial attacks. This enables efficient and reliable deployment of DL-based AMC on edge devices. We first propose two optimized models using knowledge distillation and network pruning, followed by a computationally efficient adversarial training process to improve the robustness. Experimental results on five white-box attacks show that the proposed optimized and adversarially trained models can achieve better robustness than the standard (unoptimized) model. The two optimized models also achieve higher accuracy on clean (unattacked) samples, which is essential for the reliability of DL-based solutions at edge applications.

\end{abstract}

 
\IEEEpeerreviewmaketitle
\vspace*{-1em}
\section{Introduction}
Deep learning (DL), the cornerstone of modern artificial intelligence systems, has empowered researchers to effectively solve some of the most challenging problems in diverse fields, such as healthcare, natural language processing, and computer vision \cite{deep-survey}. Inspired by their tremendous success, DL-based approaches are adopted in wireless communication domain for both classification-based \cite{vtcnn2, west2017deep,manoj2021sensing} and regression-based applications \cite{sanguinetti2019deep}. Compared to the conventional probabilistic decision theory-based methods, the DL-based approaches achieve better performance and provide significant computational advantages, such as extracting superior features directly from a large corpus of wireless signal data to develop more complex models and scalability to multiple use cases. 

In practice, to accomplish the potential of DL-based solutions for wireless communication applications, it is of paramount importance to design these solutions for successful deployment on edge devices. 
By combining the advent of the next-generation communication technology, the DL-based approaches can empower edge devices, such as drones and IoT systems, to perform intelligent wireless communication tasks efficiently and autonomously \cite{iot-sensing}. Edge devices equipped with DL capabilities can process data locally, reducing the need for transmitting large volumes of data to centralized servers, thus conserving bandwidth and enhancing privacy. However, edge devices generally have power constraints and limited computational resources, and the complex, over-parametrized DL models must be optimized before deployment to run efficiently with faster inference, less compute requirement, and lower power and memory consumption. The most commonly used methods to achieve model optimization are knowledge distillation (KD), network pruning, and model quantization \cite{mishra2020survey}. KD is a powerful method to transfer the rich knowledge learned by a complex, large deep neural network (DNN) to a lightweight network to achieve comparable performance. In network pruning, the less important neurons or weights of a DNN are identified and removed to make the network sparse. The sparsity will benefit a model to achieve faster inference and lesser storage requirements.
Model quantization also aims to achieve model storage optimization by representing the weights of a DNN at a reduced precision.

Although the ability to deploy optimized DL models on edge devices is beneficial for enhancing privacy through local data handling and computation, a critical security threat to such applications is the vulnerability of DL models to various attacks, namely, adversarial attacks, data poisoning, model extraction \cite{threat_air}, etc. This work primarily focuses on the threat towards optimized DL models against adversarial attacks, where a malicious adversary generates an adversarial example by adding a well-crafted perturbation to the input signal, which could lead to an incorrect prediction by the DL model. The adversarial perturbations are low-power signals that are hard to detect. The adversarial attacks can be classified as: a) white-box (WB) and b) black-box (BB) attacks. In a WB attack, the adversary has access to the trained DL model parameters and the training data to generate the adversarial perturbations, e.g., fast gradient method (FGM), fast gradient sign method (FGSM), projected gradient descent (PGD) \cite{adv_and_defenses}, etc. On the other hand, the adversary lacks model-related information in a BB attack, such as the universal adversarial perturbation (UAP) \cite{sadeghi2018adversarial}. 

This work focuses on the development of DL models that are optimized for edge devices as well as robust against adversarial attacks, with automatic modulation classification (AMC) of radio-frequency (RF) signals as the wireless application of interest. AMC is a safety-critical task with different applications, such as cognitive radio, signal detection and demodulation, and spectrum monitoring as well as management \cite{dobre2007survey}. In recent years, DL-based methods for AMC have been proposed based on convolutional neural networks (CNN) \cite{vtcnn2, west2017deep}, long short-term memory (LSTM) \cite{zhang2020automatic}, and transformer architectures \cite{amc-transformer}. These are generally complex networks with millions of trainable parameters that have to be optimized for deployment in edge devices. 
It has also been shown in the literature that the DL models for AMC are highly susceptible to both WB and BB adversarial attacks \cite{sadeghi2018adversarial}. Thus, several defense techniques are proposed based on adversarial training (AT) \cite{bahramali2021robust2, towards-mbr, maroto2021safeamc}, randomized smoothing \cite{towards-mbr}, GAN \cite{gan-amc-adv}, etc. Amongst the defense methods, AT is found to provide superior performance for improving the robustness of the DL models. In the AT method, adversarial examples are augmented to the training dataset so that the DL model learns features from the adversarial input space. In the literature,  the effectiveness of the AT method has been demonstrated for AMC \cite{bahramali2021robust2,maroto2021safeamc,towards-mbr}. To the best of the author's knowledge, there has not been any work in the investigation of the robustness of optimized DL models against adversarial attacks. Moreover, a key drawback of applying AT on large, complex models is the computational complexity associated with the generation of adversarial examples and the training process. The computational cost is significantly higher for examples generated using an iterative adversarial attack having a better attack success rate, such as PGD, momentum iterative method \cite{mim-2018}, etc. This results in a great challenge for performing on-device AT at the edge applications where computational resources are scarce. Thus, now more than ever, there is a great need to develop optimized, computationally efficient DL models for AMC that are also robust to adversarial attacks.

Specifically, the main contributions of this work are as follows: 
(a) We propose to implement KD and network pruning to develop the optimized and lightweight DL models for AMC. Through optimization, the computational cost of AT can also be significantly reduced to enable enhanced security of DL models with edge computing. (b) We show that both distillation and pruning are effective for developing optimized and robust DL models for AMC with a computationally efficient AT process. 
(c) For AT, we propose to utilize a combination of single-step and multi-step attacks, i.e., FGSM and PGD samples, and show that it also helps in achieving significant robustness against other unseen attacks (FGM, Deepfool, and UAP).
(d) To further ensure secure deployment, the classification accuracy of all the adversarially trained models for clean samples should not be affected significantly. We have observed that the optimized models developed in this work perform much better than the standard model after AT.
\begin{algorithm}[!t]
\small
    \caption{Distilled model}
    \label{alg:distilled-model}
    \KwIn{Student ${f}_{\mathcal{D}}(.;\boldsymbol\theta_{\mathcal{D}})$, teacher ${f}_{\mathcal{T}}(.;\boldsymbol\theta_{\mathcal{T}})$, clean training data $\mathbf{X}=\{ (\vect{x}_i, \vect{y}_i)\}_{i=1}^{N}$, $\mathcal{L}_{d}$, $\mathcal{L}_{c}$, temperature $T$, weight $\alpha$, epochs $E$, batch size $B$}
    \textbf{Initialize}: Model parameters $\boldsymbol\theta_{\mathcal{D}}$
    \hrulealg
    Number of batches per epoch: $N_{B}=\mathrm{ceil}(N/B)$
    
    \For{$m \,\, \mathrm{in} \,\, \{1,2,\dots,E\}$}{
        \For{$n \,\,\mathrm{in} \,\, \{1,2,\dots,N_B\}$}{
         Randomly sample a batch: $(\mathbf{X}_B,\mathbf{Y}_B)$

         Teacher Predictions for the batch: $\hat{\mathbf{Y}}_{\mathcal{T},B}=\mathrm{Softmax}_{T}({f}_{\mathcal{T}}(\mathbf{X}_B,\boldsymbol\theta_{\mathcal{T}})) $

         Student Predictions for the batch: $\hat{\mathbf{Y}}_{\mathcal{D},B}=\mathrm{Softmax}_{T}({f}_{\mathcal{D}}(\mathbf{X}_B,\boldsymbol\theta_{\mathcal{D}})) $

         Distillation loss: $\mathcal{L}_{d} = KL_{Div.}(\hat{\mathbf{Y}}_{\mathcal{T},B}, \hat{\mathbf{Y}}_{\mathcal{D},B})$

         Classification loss: $\mathcal{L}_{c}= CE(\hat{\mathbf{Y}}_{\mathcal{D},B}, \mathbf{Y}_B)$

         Total loss: $\mathcal{L}_{t} = \alpha\cdot\mathcal{L}_{d} + (1-\alpha)\cdot\mathcal{L}_{c}$

         // Compute gradient of $\mathcal{L}_{t}$ and update $\boldsymbol\theta_{\mathcal{D}}$

         $\boldsymbol\theta_{\mathcal{D}} \gets \boldsymbol\theta_{\mathcal{D}}-\nabla_{\boldsymbol\theta_{\mathcal{D}}}\mathcal{L}_{t}$
        }
    }
    \hrulealg
    \textbf{Output}: Distilled model ${f}_{\mathcal{D}}(.;\boldsymbol\theta_{\mathcal{D}})$
    \end{algorithm}
\vspace*{-0.5em}
\section{optimized DL models for AMC}
Two optimized DL models for AMC are developed to investigate the robustness against adversarial attacks, and in this work, they are named \emph{distilled} and \emph{distill-pruned} models.
\subsection{Distilled model}
To develop this model, we have implemented the vanilla KD method proposed in \cite{hinton2015distilling}. KD has been widely adopted for optimizing the DNN for deployment in resource-constrained edge devices \cite{kd-beyer-2022}. The main idea is to transfer the knowledge from the learned representations of a large, complex model (teacher model) to a smaller, less complex model (student model).  The student model will provide several computational advantages for edge devices, such as faster inference and less storage and power requirements. Moreover, the student model can also achieve performance comparable to or even better than the teacher model, which is an additional important benefit of KD. Algorithm \ref{alg:distilled-model} presents the development of the distilled model using the vanilla KD method \cite{hinton2015distilling}. In the algorithm, the pre-trained teacher model is denoted by ${f}_{\mathcal{T}}(.;\boldsymbol\theta_{\mathcal{T}})$, where $\boldsymbol\theta_{\mathcal{T}}$ are the trained parameters and the untrained student model is denoted by ${f}_{\mathcal{D}}(.;\boldsymbol\theta_{\mathcal{D}})$, where $\boldsymbol\theta_{\mathcal{D}}$ are trainable parameters. The trained student model obtained after the distillation process is called the distilled model. The knowledge transfer process takes place by minimizing the distance between the output probability distributions of the teacher and student models, i.e.,  $\hat{\mathbf{Y}}_{\mathcal{T},B}$ and $\hat{\mathbf{Y}}_{\mathcal{D},B}$, respectively. The output probabilities are computed by applying $\mathrm{Softmax}_T(\cdot)$ on the logit values, where the temperature parameter $T$ regulates the softness of the output probability distributions. During distillation, a high value of $T$ is used, and $T$ is set to $1$ during inference. In Algorithm \ref{alg:distilled-model}, we have used the Kullback-Leibler (KL) divergence, $KL_{Div.}(\cdot)$ to compute the distillation loss, $\mathcal{L}_{d}$ between $\hat{\mathbf{Y}}_{\mathcal{T},B}$ and $\hat{\mathbf{Y}}_{\mathcal{D},B}$. For the classification loss of the student, i.e., $\mathcal{L}_{c}$, we have used the cross-entropy loss.  The total loss for the student model, $\mathcal{L}_{t}$, is the weighted sum of $\mathcal{L}_{d}$ and $\mathcal{L}_{c}$, with  $\alpha $ being the weight of the distillation loss. 

For demonstration purposes, this work considers the VTCNN2 model in \cite{vtcnn2} as the student model and the InceptionNet model in \cite{west2017deep} as the teacher model. The VTCNN2 and the InceptionNet models have $2.83$M and $10.07$M parameters, respectively.  Both architectures are publicly available, which is beneficial for the reproducibility of our results. The parameters $T=10$ and $\alpha=0.1$ are chosen for this work. For the remainder of the manuscript, ${f}_{\mathcal{D}}$ will refer to the distilled VTCNN2 model, where knowledge is distilled from the InceptionNet teacher model. Also, the original VTCNN2 model trained specifically for AMC as in \cite{vtcnn2} (without KD) is referred to as the standard model ${f}_{\mathcal{S}}$ in this work.
\begin{algorithm}[!t]
\small 
    \caption{Distill-pruned model}
    \label{alg:distilled-pruned-model}
    \KwIn{Distilled model ${f}_{\mathcal{D}}(.;\boldsymbol\theta_{\mathcal{D}})$, normalized weight matrices $\mathbf{W}_1,\mathbf{W}_2, \dots, \mathbf{W}_L$, $L$ layers, prune-layer index $k$, data matrix $\mathbf{U}\subset \mathbf{X}$, $\eta >0$ }
    \hrulealg
    // Calculate layer-wise activations with original weights
 
    $\mathbf{Y}_{0}= \mathbf{U}$   \quad\quad\quad\quad // Input data
    
    \For{$l\,\,\mathrm{in}\,\,\{1,2,\dots,L\}$}{
    
            $\mathbf{Y}_{l} \gets \mathrm{max}(\mathbf{W}_l^\top \mathbf{Y}_{l-1}, \mathbf{0})$ // Activations before pruning
    }   
    
    $\hat{\mathbf{W}}_{k} \gets \mathrm{TRIM}(\mathbf{Y}_{k-1}, \mathbf{Y}_{k}, \mathbf{0}, \eta)$     // Apply $\mathrm{TRIM}$ on layer $k$

    Update: $\mathbf{W}_{k} \gets  \hat{\mathbf{W}}_{k}$
    \hrulealg
    \textbf{Output}: Distill-pruned model ${f}_{\mathcal{P}}(.;\boldsymbol\theta_{\mathcal{P}})$
\end{algorithm}
\subsection{Distill-pruned model}
In this method, we combine KD and network pruning to optimize the DL model for AMC further. Specifically, the goal is to incorporate the complementary benefits of the two methods, i.e., knowledge transfer from KD and sparsity from pruning, to optimize a DL model. Network pruning is powerful for model optimization because many parameters do not contribute significantly to the network's performance and, therefore, can be removed or pruned. From an edge application perspective, pruning can have several benefits, such as faster inference, less storage and compute requirements, and increased generalization.
We first obtain the distilled model ${f}_{\mathcal{D}}(.;\boldsymbol\theta_{\mathcal{D}})$ using Algorithm \ref{alg:distilled-model}, followed by applying the Net-Trim (NT) pruning method in \cite{aghasi2017net}.  The NT algorithm optimizes a model by maximizing the sparsity in the layer weights while minimizing the difference between the post-pruning output response and the initial output response of a layer. This can be formulated as a constrained optimization problem for a layer with index $k$ as given by \cite{aghasi2017net}, 
\begin{eqnarray} \label{eq:nt-optimization}
   && \hat{\mathbf{W}}_{k}= \underset{\mathbf{V}_{k}}{\arg\min} \|\mathbf{V}_{k}\|_1 \,\,
    \text{s.t.} \quad  \|\hat{\mathbf{Y}}_k- \mathbf{Y}_{k}  \|_{F} \leq \eta\, ,   \nonumber
\end{eqnarray}
where $\hat{\mathbf{Y}}_k= \mathrm{max}(\mathbf{V}_{k}^\top \mathbf{Y}_{k-1}, 0)$ is the output activation of the $k^{th}$ layer for the intermediate weight matrix $\mathbf{V}_{k}$  during optimization, with $\mathbf{Y}_{k-1}$ being the output of the previous layer, $(\cdot)^\top$ denotes transpose, $\|\cdot\|_F$ is the Frobenius norm, and $\eta >0$ is the threshold. 
In our implementation, the first fully-connected (FC) layer of the distilled VTCNN2 model ${f}_{\mathcal{D}}$ is pruned as it has the highest number of parameters, i.e., $2.7$M out of the total $ 2.83$M parameters, and pruning this layer will effectively optimize the overall model. The final sparse weight matrix $\hat{\mathbf{W}}_{k}$ is obtained by solving the problem in (\ref{eq:nt-optimization}) using the alternating direction method of multipliers (ADMM) technique \cite{aghasi2017net}. After updating the original weights 
$\mathbf{W}_{k}$ to the sparse weights $\hat{\mathbf{W}}_{k}$, we can obtain the distill-pruned model, denoted as ${f}_{\mathcal{P}}(.;\boldsymbol\theta_{\mathcal{P}})$, where $\boldsymbol\theta_{\mathcal{P}}$ are the parameters after pruning. The overall procedure is presented in Algorithm \ref{alg:distilled-pruned-model}.  
In the algorithm, the $\mathrm{TRIM}$ method depicts the iterative solution of the optimization problem in (\ref{eq:nt-optimization}) to obtain $\hat{\mathbf{W}}_{k}$ and the parameter $\eta$ controls the extent of sparsity in the weight matrix. To generate the layer-wise output activations from the distilled model, only a small subset of samples $\mathbf{U}$, randomly chosen from the training dataset $\mathbf{X}$ for ${f}_{\mathcal{D}}$, is utilized in Algorithm \ref{alg:distilled-pruned-model}. In this work, we have used $\eta=0.8$ to achieve a $96.5\%$ sparsity, which means that only $94.47$K weights are non-zero out of the total $2.7$M weights.
\begin{figure*}[!t]
	\centering
	\includegraphics[width=0.75\textwidth, height=3.2in]{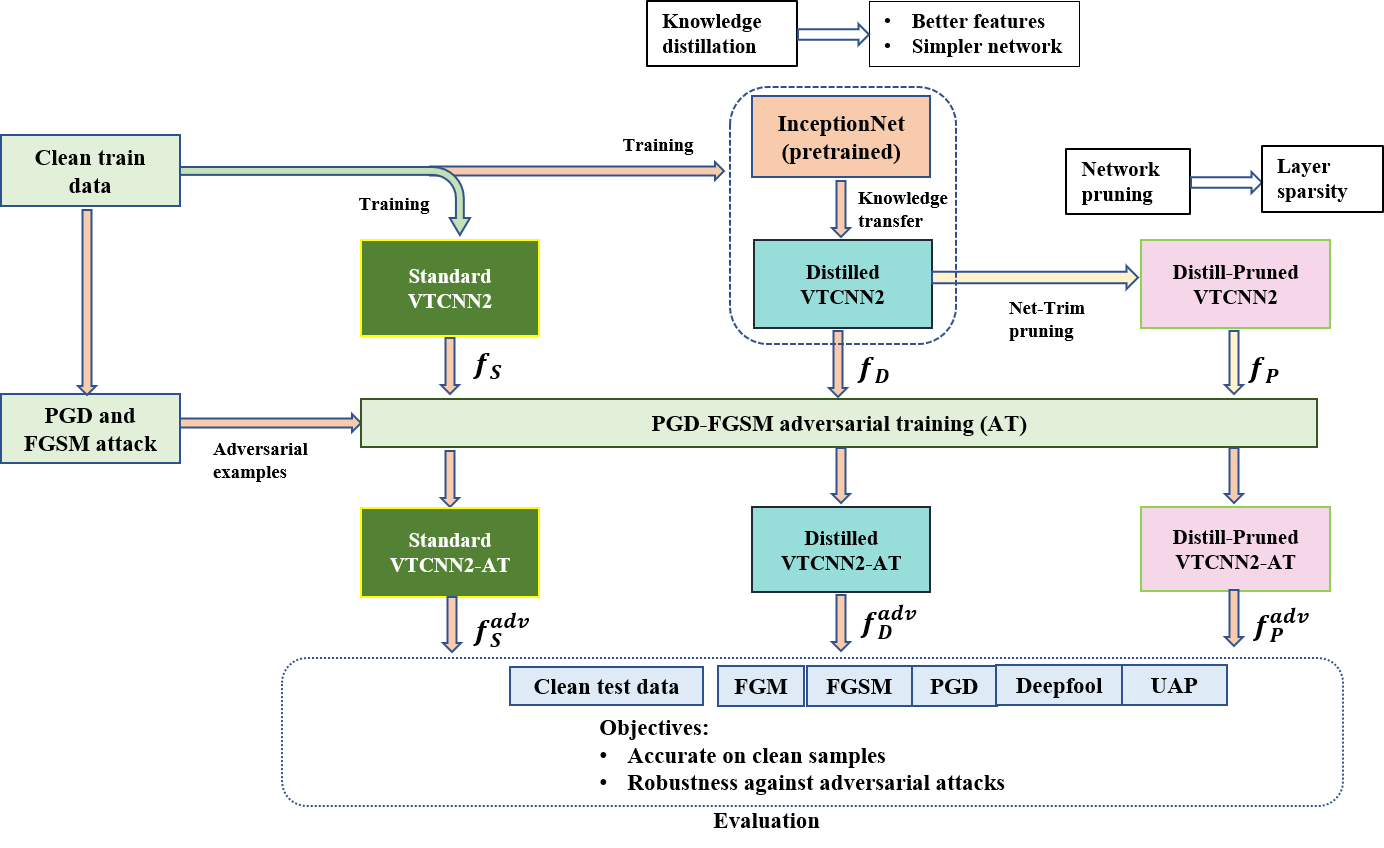}\vspace{-0.5em}
	\caption{Taxonomy for the evaluation of robustness against adversarial attacks of the proposed optimized models. }
	\label{fig:taxonomy}
    \vspace{-0.5em}
\end{figure*}
\section{Adversarial attacks}
\subsection{Attack model}
In general, we denote a trained DL-based wireless signal classifier as ${f}(.;\boldsymbol\theta): {\vect{x}} \in {\mathcal{X}} \rightarrow {\vect{y}} \in {\mathcal{Y}}$, where $\boldsymbol \theta$ are the trained parameters of the model, ${\vect{x}}$ is the clean complex-valued input RF signal (no attack) in ${\mathcal{X}} \subset {\mathbb{R}}^{2 \times n}$, referring to the in-phase ($\mathrm{I}$) and quadrature ($\mathrm{Q}$) components of dimension $n$. ${\vect{y}}$ is the clean output probability vector in ${\mathcal{Y}} \subset {\mathbb{R}}^{K}$, where $K$ is the output dimension which corresponds to the number of modulation schemes. The goal of the adversary is to generate an adversarial perturbation for the input signal $\vect{x}$, denoted as $\vect{\delta}$, specific to the attacked classifier ${f}(.;\boldsymbol \theta)$. The adversarial example is then generated as  $\vect{x}_{adv}=\vect{x}+\vect{\delta}$. When $\vect{x}_{adv}$ is provided as the input signal during inference, the classifier predicts the corresponding output label as $\hat{l}({\vect{x}_{adv}}) = \arg\max_{j} {{f}^{j}({\vect{x}_{adv}};{\boldsymbol \theta})}$, where ${{f}^{j}({\vect{x}_{adv}};{\boldsymbol \theta})}$ is the output probability of the classifier corresponding to the $j$-th class. If $l(\vect{x})$ is the original label of the clean RF signal $\vect{x}$, then the adversarial attack will be successful if $l(\vect{x}) \neq \hat{l}({\vect{x}_{adv}})$. In this work, we focus on untargeted adversarial attacks, where the predicted label $\hat{l}({\vect{x}_{adv}})$ can be any other class except the original label $l(\vect{x})$. 
\vspace{-0.5em}
\subsection{White-box attacks}
\paragraph{FGM} In this attack method, the $\vect{x}_{adv}$ is generated by solving a constrained optimization problem as given by
\begin{equation} \label{eq:fgm}
    \arg\max_{\vect{x}_{adv}}\, \mathcal{L}(\boldsymbol \theta, \vect{x}_{adv}, \vect{y}), \quad \mathrm{s.t.} \quad \norm{\vect{x}_{adv}-\vect{x}}_2 \leq \epsilon\, ,
\end{equation}
where $\mathcal{L}(\boldsymbol \theta, \vect{x}_{adv}, \vect{y})$ is the loss for ${f}(.;\boldsymbol \theta)$ when $\vect{x}_{adv}$ is the input signal, $\norm{\cdot}_2$ is the $L_2$-norm and $\epsilon$ is the allowed perturbation. The solution to (\ref{eq:fgm}) is given by
\begin{equation}
    \vect{x}_{adv}=\vect{x}+\epsilon \cdot (\norm{\nabla_{\vect{x}} \mathcal{L}(\boldsymbol \theta, \vect{x}, \vect{y})})^{-1} \mathcal{L}(\boldsymbol \theta, \vect{x}, \vect{y}) \, , 
\end{equation}
\paragraph{FGSM} This method follows the same optimization problem in (\ref{eq:fgm}) to generate the $\vect{\delta}$, except the constraint that is being subjected to is the $L_\infty$-norm instead of the $L_2$-norm, i.e., $\norm{\vect{\delta}}_\infty \leq \epsilon$. The resultant $\vect{x}_{adv}$ is given by
\begin{equation}
    \vect{x}_{adv}=\vect{x}+ \epsilon \cdot \mathrm{sign}(\nabla_{\vect{x}} \mathcal{L}(\boldsymbol \theta, \vect{x}, \vect{y}))\, , 
\end{equation}
\paragraph{PGD} This is an advanced and iterative method, which involves refining the $\vect{x}_{adv}$ at each iteration by adjusting the perturbation in the direction that maximizes the loss $\mathcal{L}(\cdot)$ while staying within the $\epsilon$-neighbourhood of the clean signal. The mathematical formulation is given by
\begin{eqnarray}
&& \vect{x}_0 = \vect{x} \label{eq:pgd1} \\ 
&&\vect{x}_{i+1}=\mathrm{clip}_{[\vect{x}, \epsilon]}\{ \vect{x}_{i} + \beta\cdot\mathrm{sign}(\nabla_{\vect{x}_{i}}\mathcal{L}(\boldsymbol \theta, \vect{x}_{i}, \vect{y}) \} \label{eq:pgd2} \\
&&\vect{x}_{adv}=\vect{x}_T \label{eq:pgd3} \, ,
\end{eqnarray}
where $\beta$ is a step size, $T$ is the number of iterations and $\mathrm{clip}_{[{\vect{x}},\epsilon]}\{{\vect{x}}_i\}$ denotes constraining the intermediate sample $\vect{x}_{i}$ in the range $[\vect{x}_{i} -\epsilon, \vect{x}_{i}+ \epsilon]$.
\paragraph{Deepfool} This is an iterative attack originally developed for binary classifiers. It is based on the idea that the minimum perturbation required for the misclassification of an input sample will be the orthogonal projection of the sample onto the decision boundary. The iterative optimization problem to generate the minimum perturbation, $\boldsymbol{\delta}_{i}$
is given by
\begin{equation} \label{eq:deepfool}
\arg\min_{{\boldsymbol{\delta}}_{i}}{||{\boldsymbol{\delta}_{i}}||_{2}}, \quad \mathrm{s.t.} \quad f({\vect{x}}_{i}) + \nabla_{\vect{x}_i} f({\vect{x}}_{i})^\mathrm{T}{\boldsymbol{\delta}}_{i} = 0,
\end{equation}

\paragraph{UAP} UAP is a method to generate adversarial perturbations that are input-agnostic and do not depend on the knowledge of the DL model; thus, these perturbations are universal in nature. In this work, we have used the PCA-based UAP method proposed in \cite{sadeghi2018adversarial}, as it is computationally efficient.
\vspace*{-0.5em}
\section{PGD-FGSM adversarial training}

Adversarial examples are generated by maximizing the loss function of a DL model as formulated in (\ref{eq:fgm}). Therefore, AT exposes the DL model to these samples by augmenting the clean training data and then aim to minimize the classification loss through standard training, which will enhance the robustness of the model against adversarial attacks.
A key objective of this work is to reduce the cost of AT so that it can be performed locally on an edge device, which will enhance the privacy and security of the application on demand.
However, most of the existing work considers incorporating adversarial samples generated from iterative or multi-step attacks as they have a high attack success rate, but it also increases the computational cost of AT significantly. In contrast, we consider incorporating examples from both single-step and multi-step attacks. This will help in reducing the computational complexity of AT because the cost of generating $N$ examples from a multi-step attack is significantly higher than the combined cost of generating $N_1$ examples from the multi-step attack and $N_2$ examples from another single-step attack, where $N_1+N_2=N $ and $N_1=N_2$.
Also, incorporating adversarial examples generated in different gradient directions can effectively improve the robustness of the models. Therefore, this work utilizes both PGD and FGSM attacks to generate adversarial examples during the AT process.
To achieve more computational efficiency, we have also fixed the weights of the first FC layer of the standard model ${f}_{\mathcal{S}}$, the distilled model ${f}_{\mathcal{D}}$, and the distill-pruned model ${f}_{\mathcal{P}}$, while performing AT.
This reduces the number of trainable parameters to around $126$K compared to the original $2.83$M parameters. The corresponding adversarially trained models are denoted as ${f}_{\mathcal{S}}^{adv}$, ${f}_{\mathcal{D}}^{adv}$, and ${f}_{\mathcal{P}}^{adv}$. Fig. \ref{fig:taxonomy} shows the taxonomy for developing the proposed robust, optimized models and the evaluation against adversarial attacks.

\begin{figure*}[!]
     \centering
     \begin{subfigure}[h]{0.32\textwidth}
         \centering
         \centerline{\includegraphics[width=5.4cm, height=3.8cm]{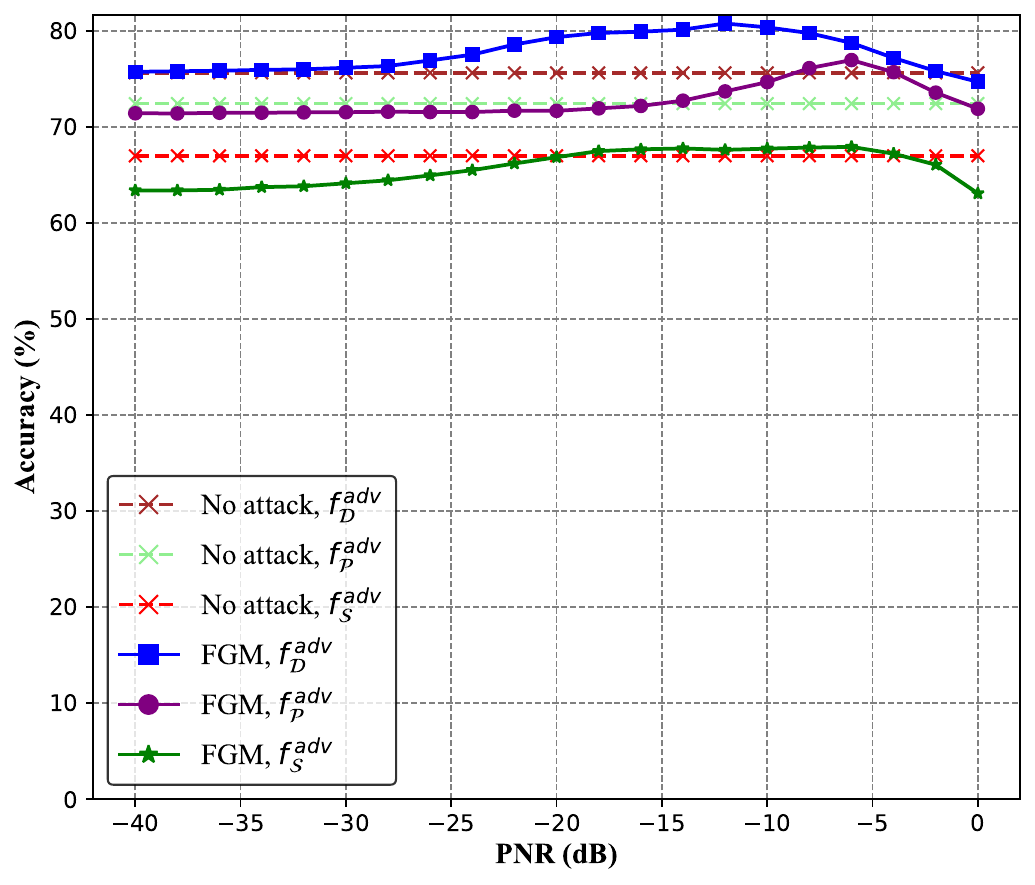}}
         \caption{FGM}
         
         \label{fig:fgm}
     \end{subfigure}
     \begin{subfigure}[h]{0.32\textwidth}
         \centering
         \centerline{\includegraphics[width=5.4cm, height=3.8cm]{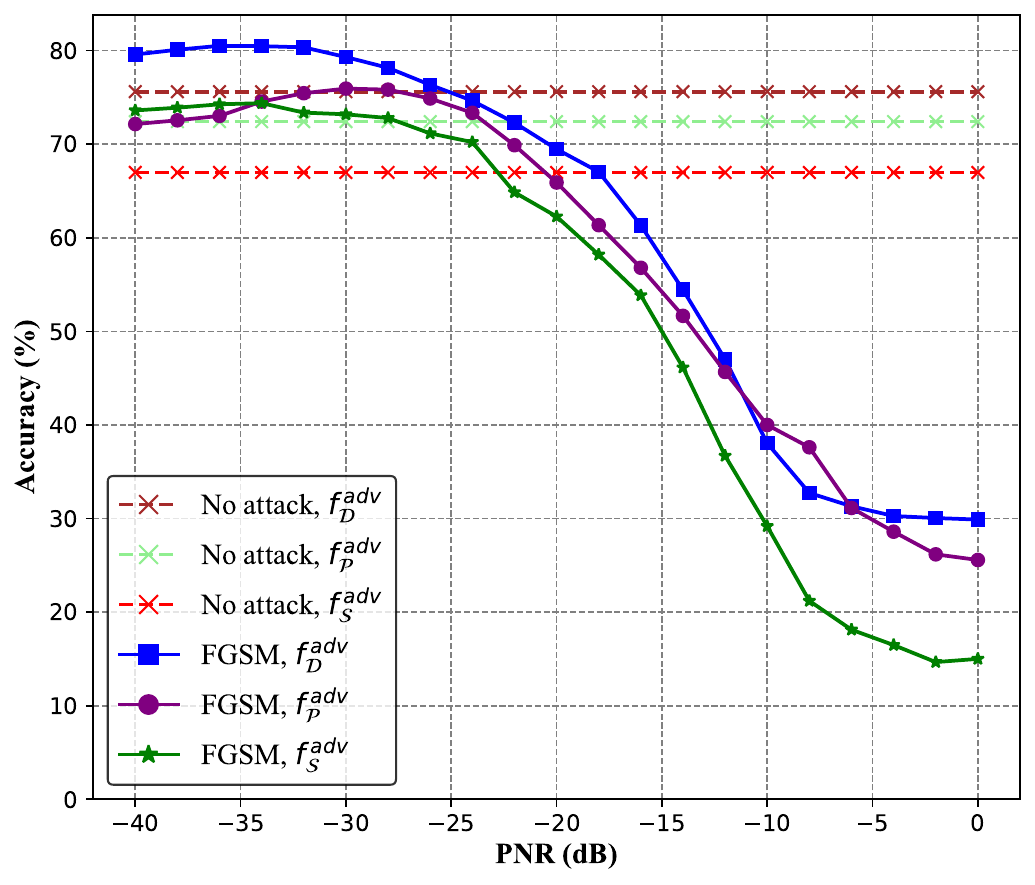}}
         \caption{FGSM}
          
         \label{fig:fgsm}
     \end{subfigure}
     \begin{subfigure}[h]{0.32\textwidth}
         \centering
         \centerline{\includegraphics[width=5.4cm, height=3.8cm]{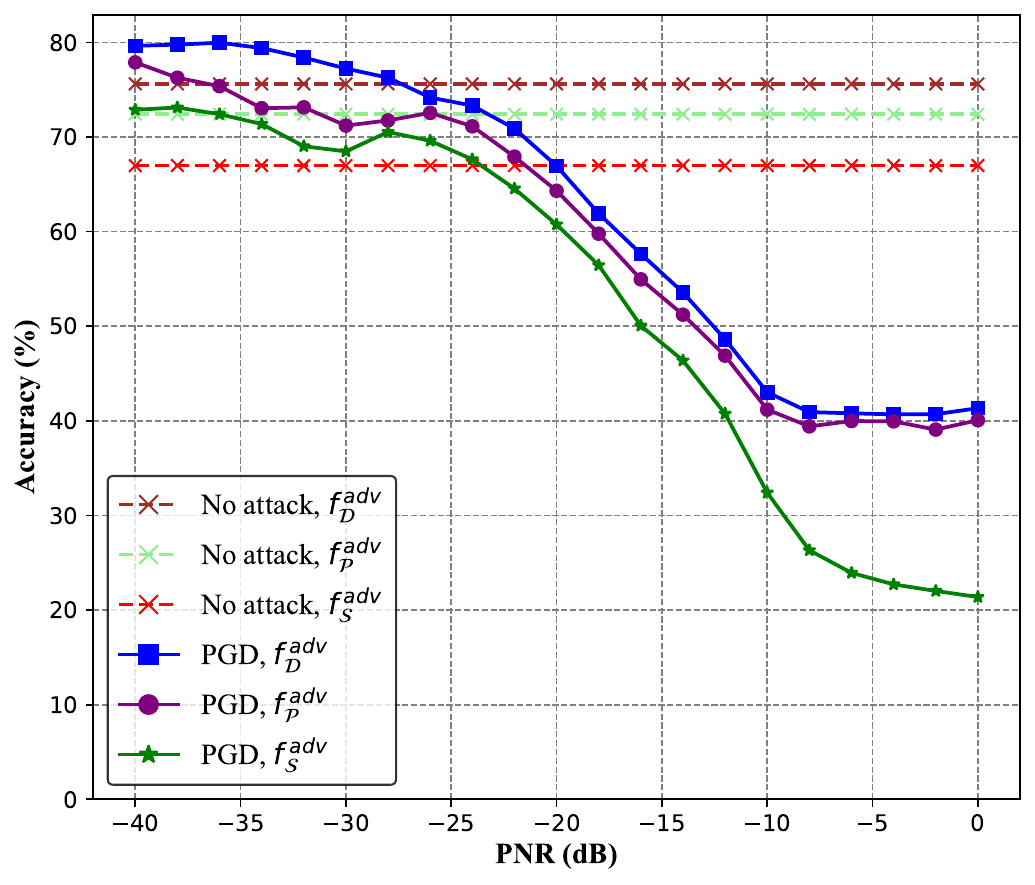}}
         \caption{PGD}
         
         \label{fig:pgd}
     \end{subfigure}

     \begin{subfigure}[h]{0.32\textwidth}
         \centering
         \centerline{\includegraphics[width=5.4cm, height=3.8cm]{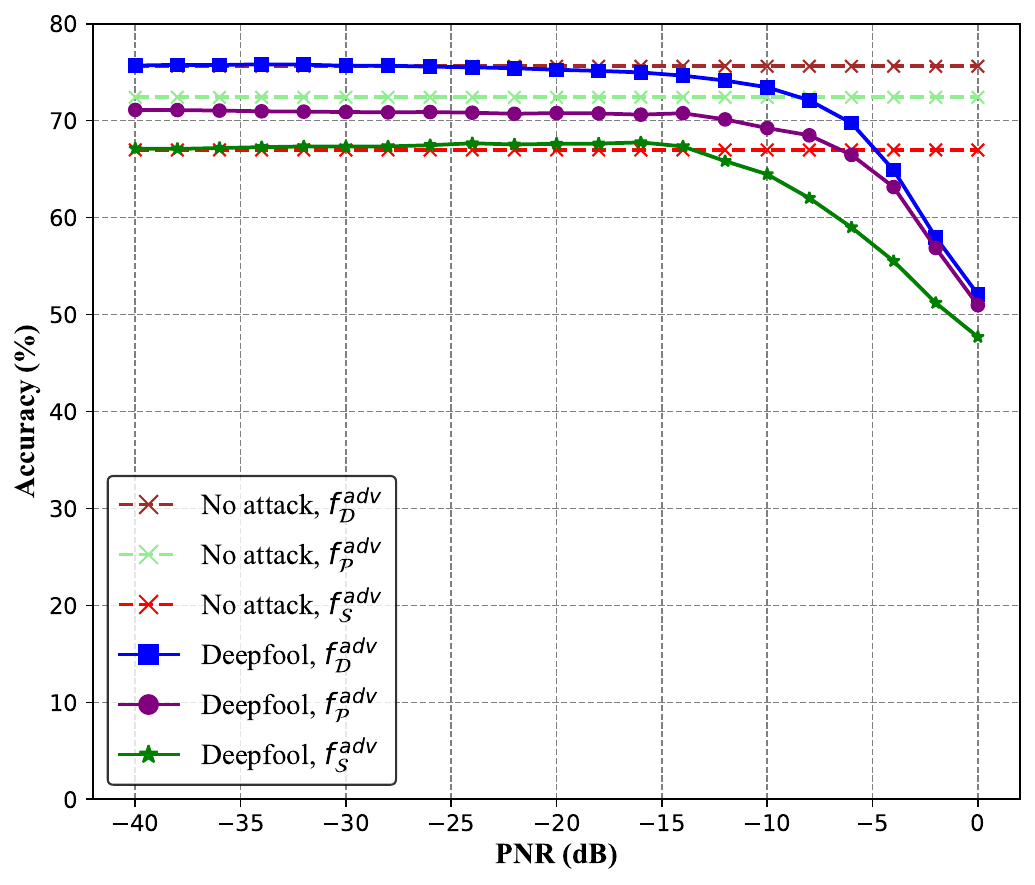}}
         \caption{Deepfool}
         \vspace*{-0.3em}
         \label{fig:deepfool}
     \end{subfigure}
     \begin{subfigure}[h]{0.32\textwidth}
         \centering
         \centerline{\includegraphics[width=5.4cm, height=3.8cm]{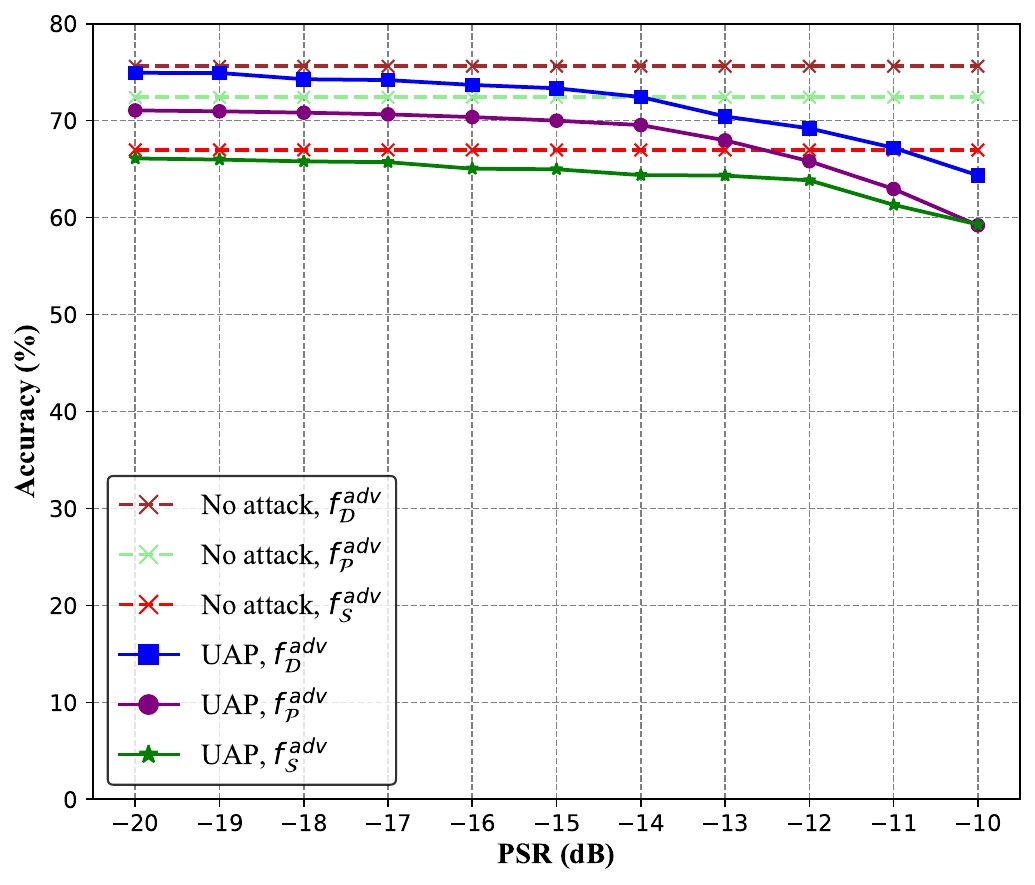}}
         \caption{UAP}
         \vspace*{-0.3em}
         \label{fig:uap-wb}
     \end{subfigure}
        \caption{\small Classification accuracy of the adversarially trained standard and optimized models for adversarial attacks at SNR=$10$ dB.} \vspace*{-0.2em}
        \label{fig:acc-attack-all}
\end{figure*}
\section{Results and discussion}

In this work, for the demonstration purpose of the proposed techniques, we have considered the well-known RML2016.10A RF modulation classification dataset \cite{vtcnn2}. The dataset consists of $220,000$ RF signals from $11$ modulation schemes. For each modulation, the signals are generated using signal-to-noise ratios (SNRs) in the range of $-20$ dB to $18$ dB with a step size of $2$ dB. Each complex-valued signal is of dimension $2\times128$, where both the $\mathrm{I}$ and $\mathrm{Q}$ components contain $128$ samples each. We have chosen this dataset because it is publicly available, which enables the reproducibility of our results. We have considered $50\%$ of the dataset as the training set, denoted as $\mathcal{D}_{Train}$, which is used to develop the models ${f}_{\mathcal{S}}$, ${f}_{\mathcal{D}}$, and ${f}_{\mathcal{P}}$. The remaining $50\%$, denoted as $\mathcal{D}_{Test}$, is used to evaluate the proposed defense method. The AT process is performed on the three models using the examples generated from $\mathcal{D}_{Train}$. The performance of the models ${f}_{\mathcal{S}}^{adv}$, ${f}_{\mathcal{D}}^{adv}$, and ${f}_{\mathcal{P}}^{adv}$ are evaluated in terms of i) robustness performance for adversarial test samples generated from five WB attacks, namely FGM, PGD, FGSM, Deepfool, and UAP, ii) classification performance of the models on the clean test samples.
Further, we define two quantities in the context of adversarial attacks: the perturbation-to-noise ratio (PNR) and the perturbation-to-signal ratio (PSR). PNR is the relation between the perturbation power and the noise power defined as $\mathrm{PNR} = \epsilon^2 \, \frac{(\mathrm{SNR}+1)}{||{\vect{x}}||_{2}^{2}}$, where $||{\vect{x}}||_{2}^{2}$ is the signal power and $\epsilon$ is the maximum allowed perturbation. PSR is the relation between the perturbation power and the signal power defined as $\mathrm{PSR} = {\mathrm{PNR}}/{\mathrm{SNR}}$.

\paragraph{Robustness against WB attacks} In Fig. \ref{fig:acc-attack-all}, we have compared the robustness performance of ${f}_{\mathcal{D}}^{adv}$, ${f}_{\mathcal{P}}^{adv}$, and ${f}_{\mathcal{S}}^{adv}$ when tested against the five representative adversarial attacks, at a fixed SNR=$10$ dB. Specifically, we have evaluated the proposed models against both single-step attacks, i.e., FGM  FGSM, and UAP, as well as multi-step (iterative) attacks, i.e. PGD and Deepfool. It can be observed that ${f}_{\mathcal{D}}^{adv}$ and ${f}_{\mathcal{P}}^{adv}$ overall performs better than the ${f}_{\mathcal{S}}^{adv}$ across all the attacks. Fig. \ref{fig:fgm} shows that for FGM attack, ${f}_{\mathcal{D}}^{adv}$ performs significantly better than ${f}_{\mathcal{S}}^{adv}$, with an average accuracy gain of $12\%$ across all PNRs. The accuracy of ${f}_{\mathcal{P}}^{adv}$ is comparable to ${f}_{\mathcal{D}}^{adv}$ at higher PNRs for FGM attack.  It can also be observed for FGSM and PGD attacks in Fig. \ref{fig:fgsm} and Fig. \ref{fig:pgd}, respectively, that both ${f}_{\mathcal{D}}^{adv}$ and ${f}_{\mathcal{P}}^{adv}$ performs significantly better than ${f}_{\mathcal{S}}^{adv}$ at the high PNR values. For example, both ${f}_{\mathcal{D}}^{adv}$ and ${f}_{\mathcal{P}}^{adv}$ achieve an accuracy gain of around $20\%$ compared to ${f}_{\mathcal{S}}^{adv}$ when evaluated for PGD attack at PNR = $0$ dB. Similarly, for FGSM attack, ${f}_{\mathcal{D}}^{adv}$ and ${f}_{\mathcal{P}}^{adv}$ achieve $15\%$ and $12\%$ higher accuracies, respectively, at PNR = $0$ dB. 
 
These results signify that a DL model that is optimized using KD can achieve improved adversarial robustness than the standard model when both are subjected to a computationally efficient AT process. We have also observed that pruning the distilled model can still provide better adversarial robustness than the standard model. This is beneficial for edge applications as we can achieve both high sparsity and robustness simultaneously with the distilled-pruned model. Our evaluation also proves that performing AT with the combination of a single-step attack (FGSM) and an iterative attack (PGD) can be computationally efficient as this also achieves robustness against unseen single and multi-step attacks, such as FGM, Deepfool, and UAP, without incorporating any adversarial samples during the AT process.
\begin{figure}[!]
	\centering
	\includegraphics[width=2.3in, height=1.8in]{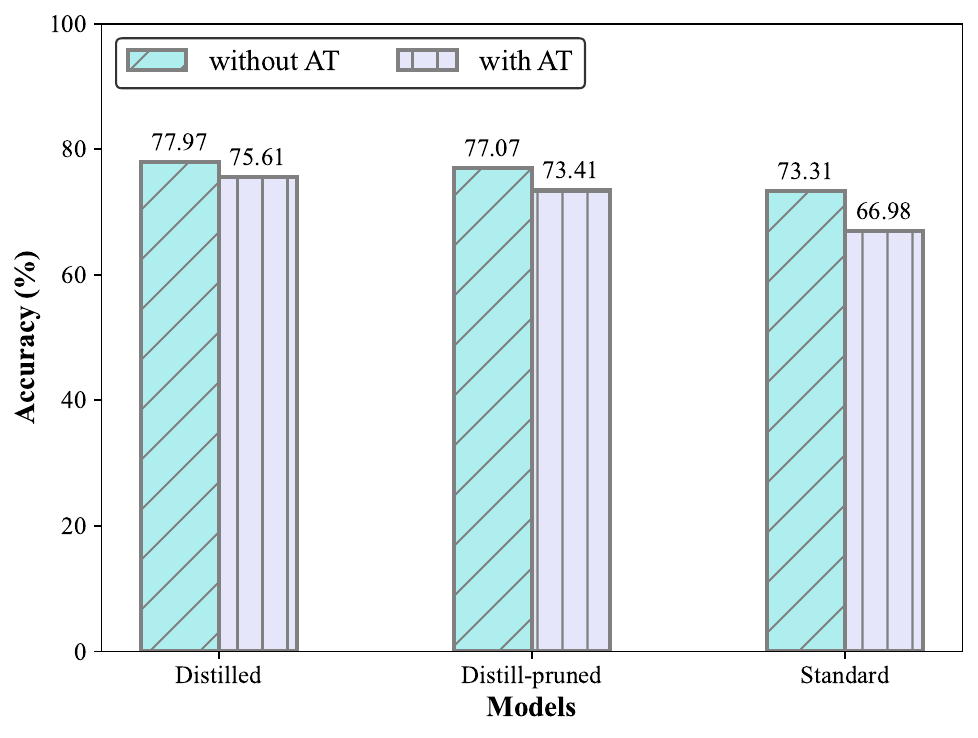}\vspace{-0.5em}
	\caption{\small Classification accuracy of the models on clean samples with and without AT at SNR=$10$ dB.}
	\label{fig:clean}
    \vspace{-0.5em}
\end{figure}
\paragraph{Performance on clean samples} We have compared the classification performance of the optimized and the standard models for the clean samples when evaluated with and without AT  at SNR=$10$ dB, as shown in Fig. \ref{fig:clean}. The AT process can lead to a reduction in accuracy for the clean samples, i.e., the clean accuracy, when compared to the model without AT. The reduction in clean accuracy affects the reliability of the model for classifying received signals without any attack. Therefore, it is important to minimize the decrease in the clean accuracy. It can be observed in Fig. \ref{fig:clean} that the drop in clean accuracy after AT is the lowest for ${f}_{\mathcal{D}}^{adv}$ and the highest for  ${f}_{\mathcal{S}}^{adv}$. Specifically, the clean accuracies of ${f}_{\mathcal{D}}$, ${f}_{\mathcal{P}}$, and ${f}_{\mathcal{S}}$ (models before AT) are $77.97\%$, $77.07\%$, and $73.31\%$, respectively. After performing AT, the accuracies of the models ${f}_{\mathcal{D}}^{adv}$, ${f}_{\mathcal{P}}^{adv}$, and ${f}_{\mathcal{S}}^{adv}$ are $75.61\%$, $73.41\%$, and $66.98\%$, respectively. It can be observed that both distilled and distill-pruned models have higher accuracies on clean samples than the standard model, before as well as after AT. This shows that KD also helps the lightweight model to learn more robust features and is not affected significantly by AT. 

\begin{figure}[t]
	\centering
	\includegraphics[width=2.3in, height=1.85in]{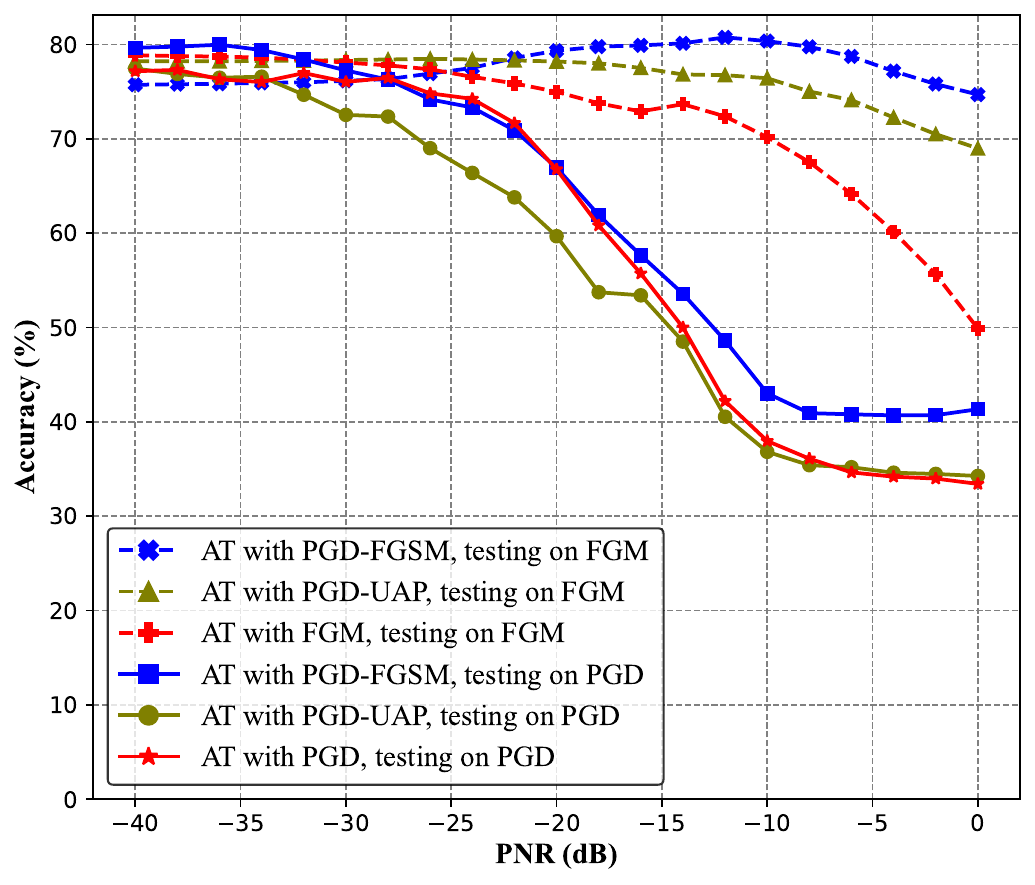}\vspace{-0.5em}
	\caption{\small Classification accuracy of ${f}_{\mathcal{D}}^{adv}$ for the PGD and FGM attacks when trained with different AT methods at SNR=$10$ dB.}
	\label{fig:at-comp}
    \vspace{-0.5em}
\end{figure}

To emphasize the choice of FGSM for AT, we have also analyzed the robustness of ${f}_{\mathcal{D}}^{adv}$ when UAP along with PGD is used for AT. UAP is also a single-step attack and, thus, can provide comparable computational benefits for AT. Fig. \ref{fig:at-comp} shows a comparison of the accuracies of ${f}_{\mathcal{D}}^{adv}$ for the PGD and FGM attacks with different AT methods at SNR=$10$ dB. For the PGD attack, it can be observed that AT with PGD and FGSM performs the best and is even better than AT with only PGD samples. Similarly, for the FGM attack, AT with PGD and FGSM achieves the highest robustness and is significantly higher than the model trained with only FGM samples. Thus, using a combination of multi and single-step attacks for AT can significantly increase the robustness of the model for FGM attack. Increasing the robustness of a DL model against the FGM attack is especially relevant for wireless communication applications as it takes into account the perturbation power ($L_2$-norm).
\vspace*{-1em}
\section{Conclusion}
In this work, we proposed two DL-based optimized models for AMC, namely distilled and distill-pruned models, based on knowledge distillation and network pruning. The primary objective of the proposed approach is to enhance the robustness of the optimized models against adversarial attacks for secure deployment in edge devices. To achieve this, we performed adversarial training with PGD and FGSM samples on the optimized models in a computationally efficient manner. Further, we investigated the robustness of these models using five adversarial attacks: FGM, FGSM, PGD, Deepfool, and UAP. Experimental results have shown that the optimized models can achieve better robustness than the standard model, with the distilled model achieving the maximum robustness across all attacks. We have demonstrated that distillation also helps with minimizing the loss in accuracy for the clean samples for the adversarially trained optimized models. Future work will incorporate developing computationally efficient, retraining-free countermeasure techniques to enable the on-device robustness improvement of DL models.

\bibliographystyle{IEEEtran}
\bibliography{bibliography}
\end{document}